\newcommand{\CLASSINPUTtoptextmargin}{19mm}%
\newcommand{\CLASSINPUTbottomtextmargin}{43mm}%
\newcommand{\CLASSINPUTinnersidemargin}{12.9mm}%
\newcommand{\CLASSINPUToutersidemargin}{12.9mm}%
\documentclass[conference,10pt,a4paper]{IEEEtran}
\usepackage{amsmath}
\usepackage{times}
\usepackage{graphicx}
\usepackage{multirow}
\usepackage[none]{hyphenat}

\usepackage{float}
\usepackage{subfig}
\usepackage{iftex}
\usepackage[backend=bibtex,defernumbers=true,sorting=none,style=ieee]{biblatex}%
\ifPDFTeX%
\usepackage{t1enc}
\usepackage{times}
\fi%
\ifXeTeX%
\usepackage{fontspec}
\fi%
\ifLuaTeX%
\usepackage{fontspec}
\fi%
\makeatletter

\def\@maketitle{\newpage
\bgroup\par\addvspace{0.5\baselineskip}\centering%
\ifCLASSOPTIONtechnote
   {\bfseries\large\@IEEEcompsoconly{\sffamily}\@title\par}\vskip 1.3em{\lineskip .5em\@IEEEcompsoconly{\sffamily}\@author
   \@IEEEspecialpapernotice\par{\@IEEEcompsoconly{\vskip 1.5em\relax
   \@IEEEtitleabstractindextextbox{\@IEEEtitleabstractindextext}\par
   \hfill\@IEEEcompsocdiamondline\hfill\hbox{}\par}}}\relax
\else
   \vskip0.2em{\EuMWtitlesize\ifCLASSOPTIONtransmag\bfseries\LARGE\fi\@IEEEcompsoconly{\sffamily}\@IEEEcompsocconfonly{\normalfont\normalsize\vskip 2\@IEEEnormalsizeunitybaselineskip
   \bfseries\Large}\@title\par}\vskip1.0em\par
   \ifCLASSOPTIONconference%
      {\@IEEEspecialpapernotice\mbox{}\vskip\@IEEEauthorblockconfadjspace%
       \mbox{}\hfill\begin{@IEEEauthorhalign}\@author\end{@IEEEauthorhalign}\hfill\mbox{}\par}\relax
   \else
      \ifCLASSOPTIONpeerreviewca
         {\@IEEEcompsoconly{\sffamily}\@IEEEspecialpapernotice\mbox{}\vskip\@IEEEauthorblockconfadjspace%
          \mbox{}\hfill\begin{@IEEEauthorhalign}\@author\end{@IEEEauthorhalign}\hfill\mbox{}\par
          {\@IEEEcompsoconly{\vskip 1.5em\relax
           \@IEEEtitleabstractindextextbox{\@IEEEtitleabstractindextext}\par\hfill
           \@IEEEcompsocdiamondline\hfill\hbox{}\par}}}\relax
      \else
         \ifCLASSOPTIONtransmag
           {\@IEEEspecialpapernotice\mbox{}\vskip\@IEEEauthorblockconfadjspace%
            \mbox{}\hfill\begin{@IEEEauthorhalign}\@author\end{@IEEEauthorhalign}\hfill\mbox{}\par
           {\vspace{0.5\baselineskip}\relax\@IEEEtitleabstractindextextbox{\@IEEEtitleabstractindextext}\vspace{-1\baselineskip}\par}}\relax
         \else
           {\lineskip.5em\@IEEEcompsoconly{\sffamily}\sublargesize\@author\@IEEEspecialpapernotice\par
           {\@IEEEcompsoconly{\vskip 1.5em\relax
            \@IEEEtitleabstractindextextbox{\@IEEEtitleabstractindextext}\par\hfill
            \@IEEEcompsocdiamondline\hfill\hbox{}\par}}}\relax
         \fi
      \fi
   \fi
\fi\par\addvspace{0.0\baselineskip}\egroup}

\def\EuMWtitlesize{\@setfontsize{\EuMWtitlesize}{24}{24pt}}
\def\EuMWauthorsize{\@setfontsize{\EuMWauthorsize}{11}{11pt}}
\def\EuMWaffilsize{\@setfontsize{\EuMWaffilsize}{10}{10pt}}
\def\EuMWcaptionsize{\@setfontsize{\EuMWcaptionsize}{9}{10pt}}
\def\EuMWbibsize{\@setfontsize{\EuMWbibsize}{8}{10pt}}

\def\@IEEEauthorblockNstyle{\EuMWauthorsize\@IEEEcompsocnotconfonly{\sffamily}\@IEEEcompsocconfonly{\large}}
\def\@IEEEauthorblockAstyle{\EuMWaffilsize\@IEEEcompsocnotconfonly{\sffamily}\@IEEEcompsocconfonly{\itshape}\@IEEEcompsocconfonly{\large}}
\def\@IEEEauthordefaulttextstyle{\EuMWauthorsize\@IEEEcompsocnotconfonly{\sffamily}\sublargesize}

\def\thebibliography#1{\section*{\refname}%
    \addcontentsline{toc}{section}{\refname}%
    \EuMWbibsize\@IEEEcompsocconfonly{\small}\vskip 0.3\baselineskip plus 0.1\baselineskip minus 0.1\baselineskip
    \list{\@biblabel{\@arabic\c@enumiv}}%
    {\settowidth\labelwidth{\@biblabel{#1}}%
    \leftmargin\labelwidth
    \advance\leftmargin\labelsep\relax
    \itemsep \IEEEbibitemsep\relax
    \usecounter{enumiv}%
    \let\p@enumiv\@empty
    \renewcommand\theenumiv{\@arabic\c@enumiv}}%
    \let\@IEEElatexbibitem\bibitem%
    \def\bibitem{\@IEEEbibitemprefix\@IEEElatexbibitem}%
\def\newblock{\hskip .11em plus .33em minus .07em}%
\ifCLASSOPTIONtechnote\sloppy\clubpenalty4000\widowpenalty4000\interlinepenalty100%
\else\sloppy\clubpenalty4000\widowpenalty4000\interlinepenalty500\fi%
    \sfcode`\.=1000\relax}

\long\def\@makecaption#1#2{%
\ifx\@captype\@IEEEtablestring%
\par\@IEEEtabletopskipstrut
\else
\@IEEEfigurecaptionsepspace
\fi
\setbox\@tempboxa\hbox{\normalfont\footnotesize {#1.}\nobreakspace\nobreakspace #2}%
\ifdim \wd\@tempboxa >\hsize%
\setbox\@tempboxa\hbox{\normalfont\footnotesize {#1.}\nobreakspace\nobreakspace}%
\parbox[t]{\hsize}{\normalfont\footnotesize\noindent\unhbox\@tempboxa#2}%
\else
\ifCLASSOPTIONconference \hbox to\hsize{\normalfont\footnotesize\hfil\box\@tempboxa\hfil}%
\else \hbox to\hsize{\normalfont\footnotesize\box\@tempboxa\hfil}%
\fi\fi
\ifx\@captype\@IEEEtablestring%
\@IEEEtablecaptionsepspace
\else
\fi}

\newlength\tablecaptiontotableskip
\newlength\figuretocaptionskip
\def\@IEEEfigurecaptionsepspace{\vskip\figuretocaptionskip\relax}%
\def\@IEEEtablecaptionsepspace{\vskip\tablecaptiontotableskip\relax}%

\def\abstract{\normalfont%
\@IEEEabskeysecsize\bfseries\textit{\abstractname}\,\bfseries\textit{---}\,%
\@IEEEgobbleleadPARNLSP}%

\def\IEEEkeywords{\normalfont%
\@IEEEabskeysecsize\bfseries\textit{\IEEEkeywordsname}\,\bfseries\textit{---}\,%
\@IEEEgobbleleadPARNLSP}%
\def\endIEEEkeywords{\relax\vspace{0.67ex}%
\par\if@twocolumn\else\endquotation\fi%
\normalsize\normalfont}%

\def\@IEEEauthorblockNtopspace{0ex}
\def\@IEEEauthorblockAtopspace{1mm}
\def\tablename{Table}
\def\thetable{\arabic{table}}
\def\IEEEkeywordsname{Keywords}
\def\subsubsection{\@startsection{subsubsection}{3}{\z@}{1.5ex plus 1.5ex minus 0.5ex}%
{0.7ex plus .5ex minus 0ex}{\normalfont\normalsize\itshape}}%
\newlength{\CPheadmatchindent}%
\def\@seccntformat#1{\hbox to\CPheadmatchindent{\csname the#1dis\endcsname}\hskip 0.1em \relax}
\IEEEilabelindentA \parindent
\IEEEilabelindent \IEEEilabelindentA
\IEEEelabelindent \parindent
\IEEEdlabelindent \parindent
\IEEElabelindent \parindent
\makeatother

\begin{document}
\raggedbottom
%
%
%
\title{Human Skin Permittivity Characterization for Mobile Handset Evaluation at Sub-THz}
%
%
\author{%
\IEEEauthorblockN{%
Bing Xue, 
Katsuyuki Haneda, 
Clemens Icheln, and
Juha Ala-Laurinaho
}
\IEEEauthorblockA{%
Department of Electronics and Nanoengineering, Aalto University, Finland\\
firstname.lastname@aalto.fi\\
}
}
%
\maketitle

\begin{abstract}
This manuscript proposes a method for characterizing the complex permittivity of the human finger skin based on an open-ended waveguide covered with a thin dielectric sheet at sub-terahertz frequencies. The measurement system is initially analyzed through full-wave simulations with a detailed finger model. Next, the model is simplified by replacing the finger with an infinite sheet of human skin to calculate the forward electromagnetic problem related to the permittivity characterization. Following this, a radial basis network is employed to train the inverse problem solver. Finally, the complex permittivities of finger skins are characterized for 10 volunteers. The variations in complex relative permittivity across different individuals and skin regions are analyzed, revealing a deviation of $<\pm 1.5$ for both the dielectric constants and loss factors across 140 to 220 GHz. Repeated measurements at the same location on the finger demonstrate good repeatability with a relative estimation uncertainty $<\pm 1.5\%$. 

\end{abstract}

\begin{IEEEkeywords}
Human skin permittivity, sub-THz, finger permittivity, radial basis network, forward problem model 
\end{IEEEkeywords}

\section{Introduction}
The sub-terahertz (sub-THz) band has garnered attention from researchers in the telecommunication community due to its potential for high data-rate characteristics~\cite{zhang2022out, nor2022effect}. Handset antennas are sensitive to the radio environment, especially in close proximity to obstructions such as fingers, hand palms, and human bodies~\cite{Krogerus2007}. To investigate human-antenna interaction for enhanced handset antenna designs, knowledge of the permittivity of human tissues becomes crucial. However, there exists a notable scarcity of research specifically addressing sub-THz frequencies, with a lack of measurement-based permittivity estimates for human tissues such as fingers, palms, and the body. The skin depth of typical human skin in millimeter wave (mmW) bands is less than $2~\rm mm$, as indicated by the formulas provided in TABLE 1.1 of~\cite{pozar2011}. Consequently, it is appropriate to model the effects of humans on antennas as a skin-antenna interaction~\cite{Xue2022}. 

Permittivity measurements have been conducted beyond 60-GHz frequencies in various studies~\cite{hosseini2016wideband, Zhu2021complex, wang2022fast, aliouane2022material}. Many of these investigations rely on the transmission and reflection of electromagnetic waves, primarily targeting layered materials. As a consequence, these methods are typically applicable only to flat slices of human skin. In practical scenarios, it becomes essential to conduct measurements on living skin to study realistic effects on mobile terminal antennas. To address this need, the open-ended probe method has been employed in 5G mmW bands for characterizing human skin permittivity in~\cite{gao2018towards, Zhekov2019}, where some important parts of human skin, including fingers, palms, arms, and so on, were measured, yielding convincing permittivities for human-antenna interaction studies. In this manuscript, we extend this open-ended probe method to sub-THz frequencies to measure human finger skin permittivity. Differing from previous publications, we introduce a simplified full-wave simulation model and employ a machine-learning approach to train a network for estimating human skin permittivities.

\section{Permittivity Characterization}
\label{sec:permmeasure}
\subsection{Human Skin Characteristics in Sub-THz Bands}
\label{sec:charact}
Several researchers have proposed a multi-layer human skin model for THz bands~\cite{zakharov2009full, sasaki2017monte, betzalel2017modeling}. Utilizing Eq.(4) in~\cite{feldman2009electromagnetic}, typical effective relative permittivities are $2.7-0.1 \rm i$ for the stratum corneum, $3.3-5.2 \rm i$ for the epidermis, and $3.9-5.2 \rm i$ for the dermis at sub-THz frequencies. As indicated by~\cite{zakharov2009full}, the thickness of the stratum corneum varies from $15~\rm \mu m$ on arms to $300~\rm \mu m$ on fingers. Consequently, the effective permittivity of human skin at sub-THz frequencies differs depending on anatomical locations and varies among individuals~\cite{zakharov2009full}.
\subsection{Existing Problems}
The WR5 ($140-220~\rm GHz$) extender of the vector network analyzer (VNA) is equipped with an open-ended waveguide featuring a flange, enabling the measurement of finger permittivities without the need for designing a new probe. In comparison to 5G mmW-band measurements~\cite{gao2018towards}, the measured reflection coefficients are more sensitive to pressure exerted by fingers on the open-ended waveguide. Increased pressure results in more parts of the finger being pressed into the open-ended waveguide, introducing a skin protrusion problem that leads to significant inaccuracies in estimating finger permittivities~\cite{gao2018towards}. To mitigate this,~\cite{gao2018towards} used copy papers to flatten the skin's surface and improve measurement accuracy in 5G mmW bands. Although copy paper between the waveguide and finger can reduce skin protrusion compared to the finger without copy paper, even a tiny protrusion can cause a significant measurement deviation if the frequency is high enough, especially beyond $100$~GHz. Consequently, the effectiveness of this approach is limited, since copy papers are soft and, therefore, easily deformable.

\subsection{Our Proposed Method}
\begin{figure}[htbp]
    \centering
    \vspace{-\baselineskip}
	\includegraphics[width=0.7\linewidth]{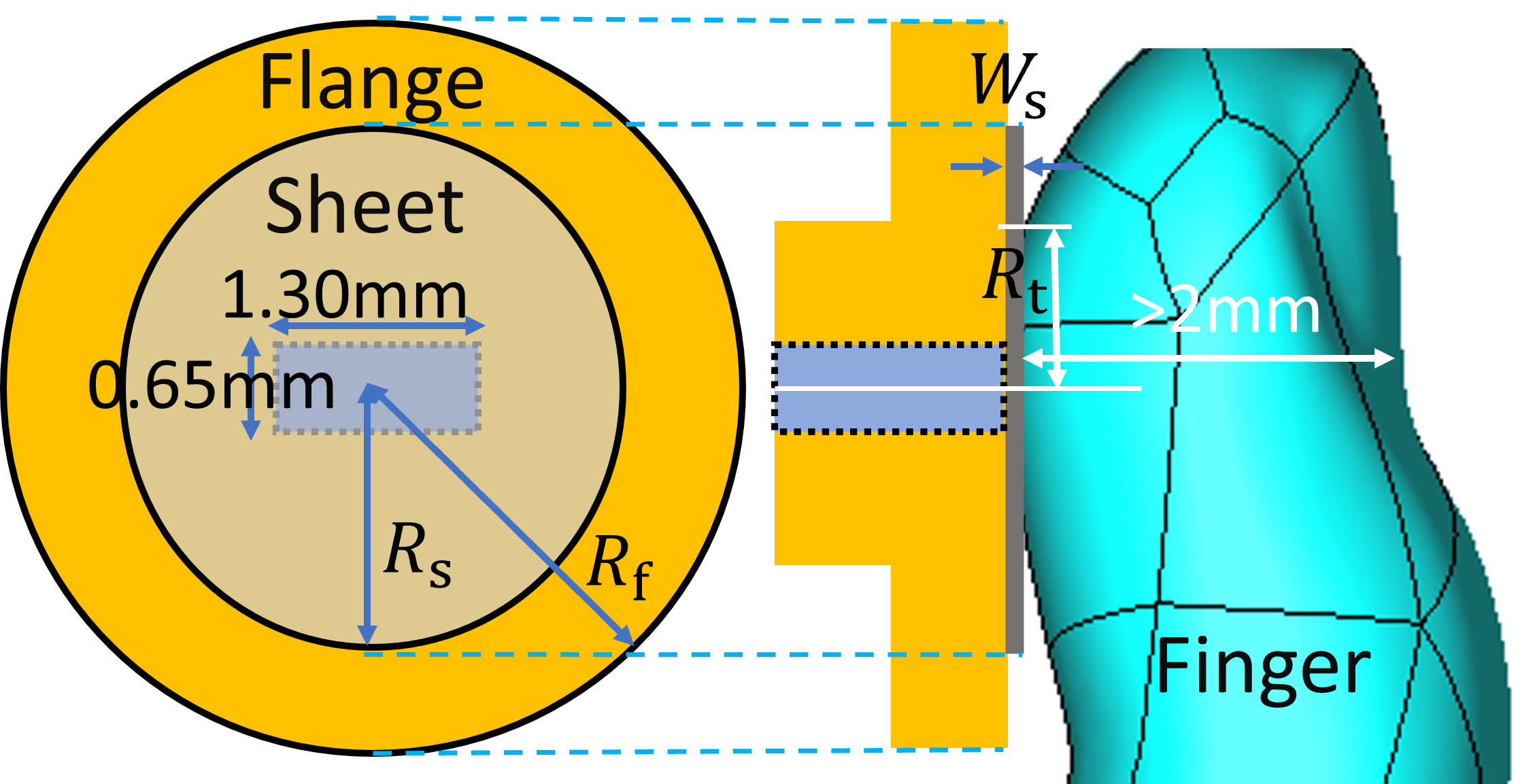}
	\caption{The front and side views of the simulation model to study the effects of the varying contact area radius. Finger skin permittivity: $\varepsilon_{\rm r} = 4.0-\frac{16.0}{\varepsilon_0 \omega}\rm i$.}
	\label{fig:simulationmodel}
\end{figure}

To address the mentioned limitations, we employ a hard and thin sheet placed on the open-ended waveguide. Following the findings in~\cite{wang2022fast}, the thickness of the sheet should be smaller than $0.1$ wavelength to reduce electromagnetic emissions along the flange plate directions and higher transmission modes of waveguides. Consequently, we opt for a hard sheet, specifically Rogers 4350B, with a relative permittivity $\varepsilon_s = 3.33-0.123\rm i$ and a thickness of $W_{\rm s} = 0.1~\rm mm$. Open-ended waveguide is an antenna that can excite waves along the flange. When the sheet is thin enough, the surface wave along the flange will be easily attenuated by the high-loss human tissues. When the contact area of the firmly pressed finger on the sheet is large enough, we can consider the surface waves to be completely attenuated. To determine the required contact area, full-wave simulations are implemented, as illustrated in Fig.~\ref{fig:simulationmodel}. The waveguide flange radius is $R_{\rm f} = 6~\rm mm$, while the sheet radius is also $R_{\rm s} = 6~\rm mm$. We assume that finger skin permittivity is $\varepsilon_{\rm r} = 4.0-\frac{16.0}{\varepsilon_0 \omega}\rm i$, following the knowledge presented in~\cite{feldman2009electromagnetic}, where $\varepsilon_0$ is the vacuum permittivity, and $\omega$ is the angular frequency. 

\begin{figure}[htbp]
    \centering
\subfloat[]{
\centering\label{fig:RtA}
     \includegraphics[width=0.5\linewidth]{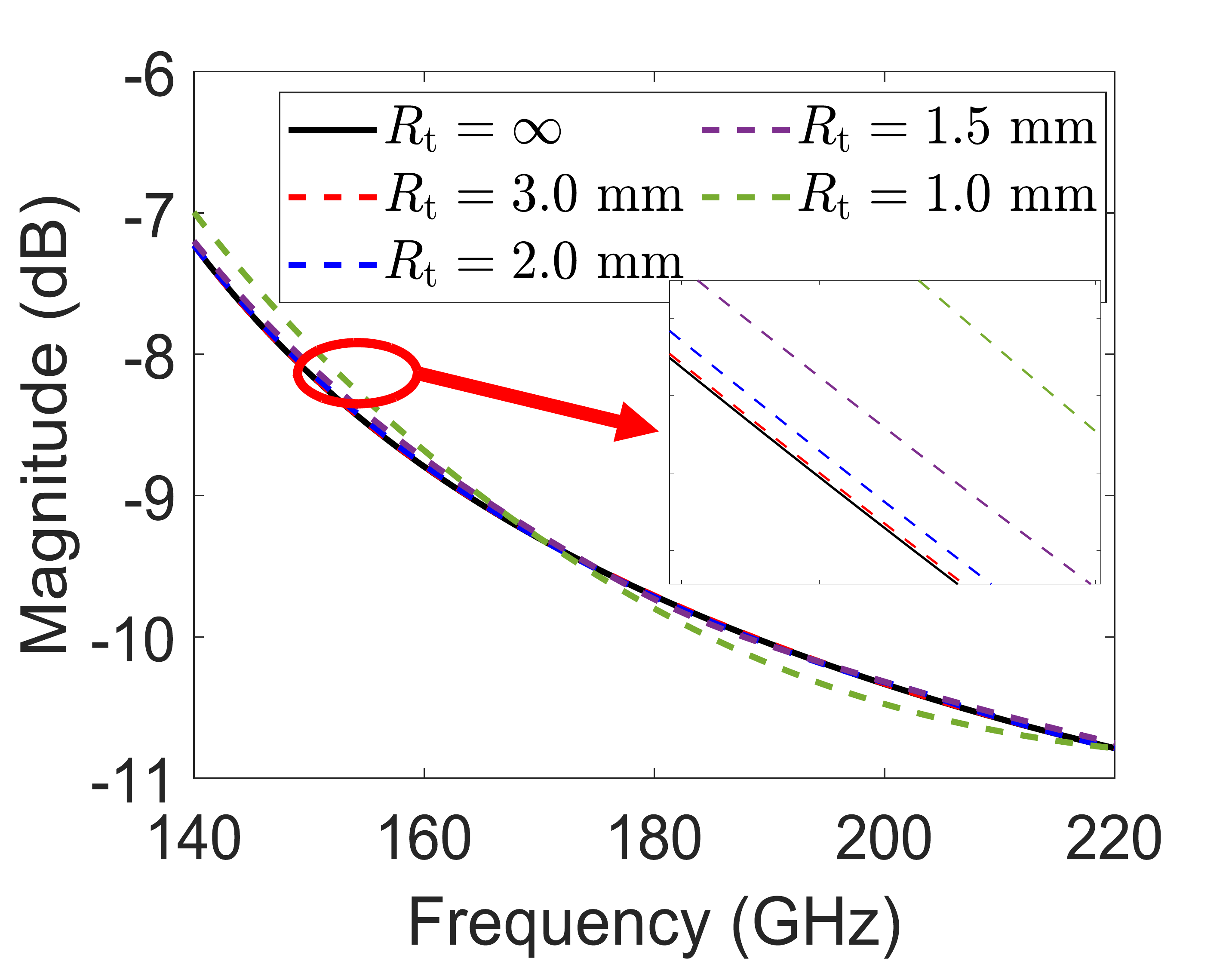}}
\subfloat[]{
\centering
	\label{fig:Rtangle}\includegraphics[width=0.5\linewidth]{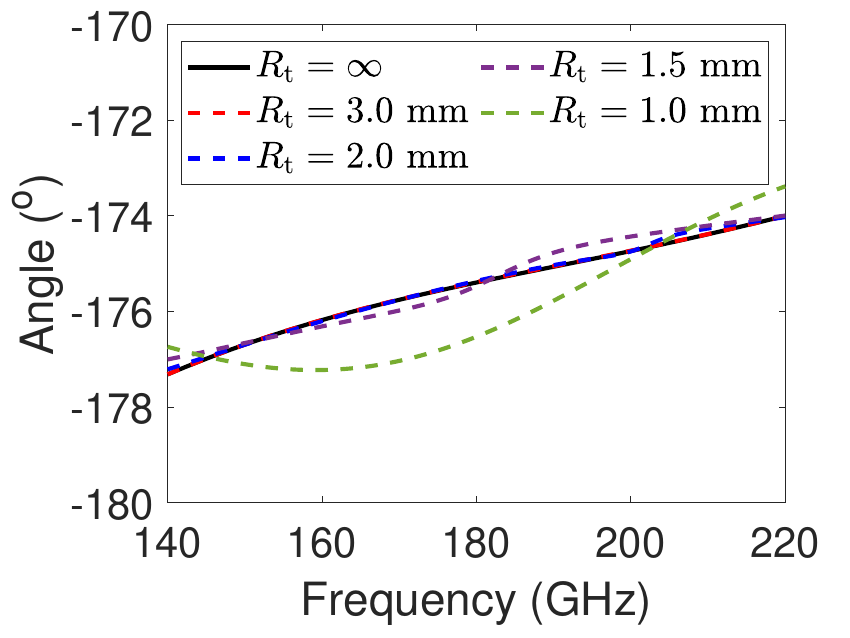}}
	\caption{(a) The magnitudes and (b) angles of reflection coefficients at the open end of the waveguide for the body part of the finger with different $R_{\rm t}$ as shown in Fig.~\ref{fig:simulationmodel}.}
	\label{fig:Rt}
 \vspace{-\baselineskip}
\end{figure}

We vary the radius of the contact area between the skin under measurement and the sheet  ($R_{\rm t}$) from $1.0~\rm mm$ to $3.0~\rm mm$, and also consider the case where $R_{\rm t} = \infty$. The magnitudes of reflection coefficients at the open end of the waveguide are depicted in Fig.~\ref{fig:RtA}. Observing the results, it becomes evident that for $R_{\rm t}>2.0~\rm mm$, the reflection coefficients closely resemble those obtained at $R_{\rm t}=\infty$. Additionally, $R_{\rm t} \geq 2.0~\rm mm$ introduces less than $0.5^\circ$ difference compared to $R_{\rm t} = \infty$ in Fig.~\ref{fig:Rtangle}. To accommodate variations in finger permittivity among different individuals, practical measurements necessitate $R_{\rm t} > 3.0~\rm mm$. This ensures that external factors like positioning pins or other structures of the open-ended waveguide, which may cause discontinuity in waves, do not significantly influence the measured reflection coefficients. This approach allows treating both the sheet and human skin as infinitely large in the context of measurements, and therefore also allows to use a simplified simulation model to produce training data for machine learning. 

When we change the human skin permittivities from 3 to 6 for the real part and from 1 to 4 for the imaginary part, which is a probable human skin permittivity range based on the knowledge in~\cite{feldman2009electromagnetic}, $R_{\rm t} > 3.0~\rm mm$ can still offer a high accuracy as $R_{\rm t} = \infty$. For brevity, they are not shown here. However, for other frequency ranges, we may need to change the size of the contact area in measurements. Therefore, we need prior knowledge of the human skin permittivity.

\begin{figure}[htbp]
 \vspace{-\baselineskip}
    \centering
	\includegraphics[width=0.65\linewidth]{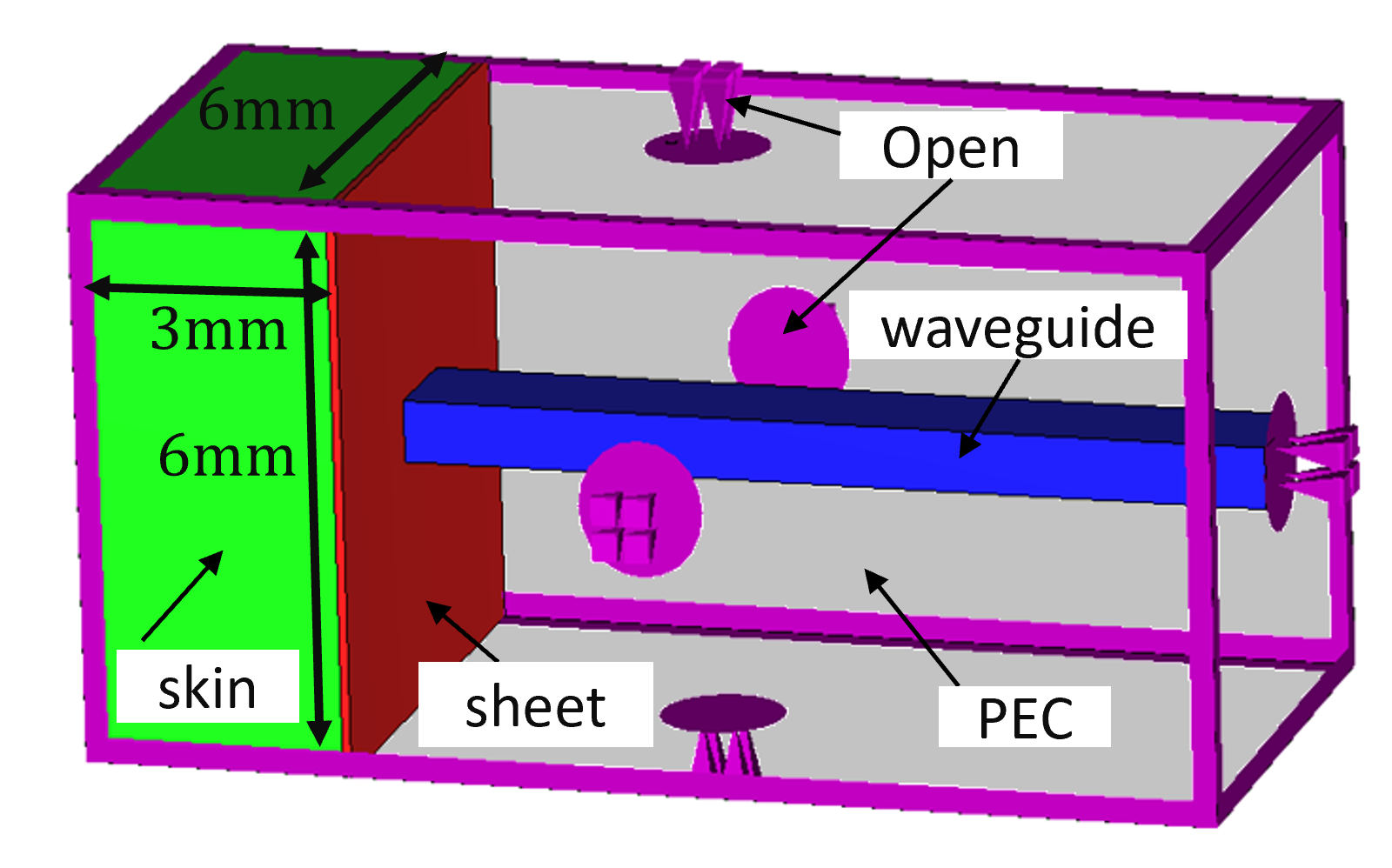}
	\caption{The calculation model of the forward problem.}
	\label{fig:forwardissue}
\end{figure}

The forward electromagnetic problem of permittivity measurements is formulated as
\begin{equation}
{\Gamma = F(\varepsilon_{\rm r})},
\label{eq:forwardproblem}
\end{equation}
where $\Gamma$ is the reflection coefficient at the open end of the waveguide, and $\varepsilon_{\rm r}$ is the complex permittivity. The simplified simulation model for the forward problem, depicted in Fig.~\ref{fig:forwardissue}, is constructed using \textit{CST Studio Suite} and solved by the frequency domain solver. In the model, the blue region represents 'air' in the waveguide, and the background material, highlighted in gray, is a perfect electric conductor (PEC). The dimensions of the skin are $6\times 6\times 3~\rm mm^3$, while the Rogers 4350B sheet measures $6\times 6\times 0.1~\rm mm^3$. The reference plane for the reflection coefficients is the open end of the WR5 waveguide, consistent with the measurement setup. The boundary of the box enclosing the entire model is set to 'open', ensuring the PEC fills the box. This setup allows the simulation model to emulate an infinite sheet and finger skin, meeting the same conditions as the measurements. To extract the reflection coefficients at the open end of the waveguide, the length of the waveguide is de-embedded. The $\rm TE_{10}$ mode is excited at the waveguide input.

To derive the inverse-problem solver, i.e., $\varepsilon_{\rm r} = F^{-1}(\Gamma)$ for permittivity characterizations, we utilize a radial basis network (RBN)\footnote{Functions such as `newrbe' in \textit{MATLAB} can be employed for this purpose.}. The input parameters of the solver consist of $\Gamma_r$ and $\Gamma_i$, representing the real and imaginary parts of the reflection coefficients. The outputs are denoted by $\varepsilon^\prime$ and $\varepsilon^{\prime\prime}$, corresponding to the real and imaginary parts of the permittivity.

Finger permittivity estimation involves the following steps:
\begin{enumerate}
    \item Calibrate the vector network analyzer (VNA) at the interface of the open-ended waveguide.
    \item Place a finger on the open end of the waveguide with the RO4350B sheet in between, and record the reflection coefficients $\Gamma$ across the frequency range.
    \item Repeat step 2) for different parts of fingers from multiple individuals.
    \item Obtain training data by simulating the setup illustrated in Fig.~\ref{fig:forwardissue} and then train the inverse-problem solver based on RBN using this data.
    \item Input the measured $\Gamma$ values into the trained inverse-problem solver to estimate the finger permittivity at each frequency.
\end{enumerate}

\section{Finger Permittivity Characterization}
\subsection{Training Data}
The spread of the radial basis network was set to $1.0$. Utilizing prior knowledge about human finger permittivity presented in Section~\ref{sec:charact}, the training data is generated by sweeping the real part of skin permittivity from $3$ to $6$, and the imaginary part from $1$ to $4$, with $1000$ uniformly distributed samples within this range. The full-wave simulation frequency varies from $140~\rm GHz$ to $220~\rm GHz$ across $1001$ uniformly sampled points.

After obtaining reflection coefficients, the dataset is split into $90\%$ training data and $10\%$ testing data. The relative mean error, defined as ${err} = \frac{\left|\varepsilon_{\rm r}-\bar{\varepsilon}_{\rm r}\right|} {\left|\varepsilon_{\rm r}\right|}$, where $\varepsilon_{\rm r}$ is the true value, and $\bar{\varepsilon}_{\rm r}$ is the estimate, of complex permittivity for the testing data is less than $0.05\%$, indicating high estimation accuracy. It is noteworthy that, to simplify the training model, the inverse problem solver is trained for each frequency point individually. On the other hand, this leads to more reliable results considering frequency-dependent permittivities which are not possible to model accurately for the training data.
\subsection{S-parameter Measurements}
\begin{figure}[htbp]
 \vspace{-\baselineskip}
    \centering
	\includegraphics[width=0.8\linewidth]{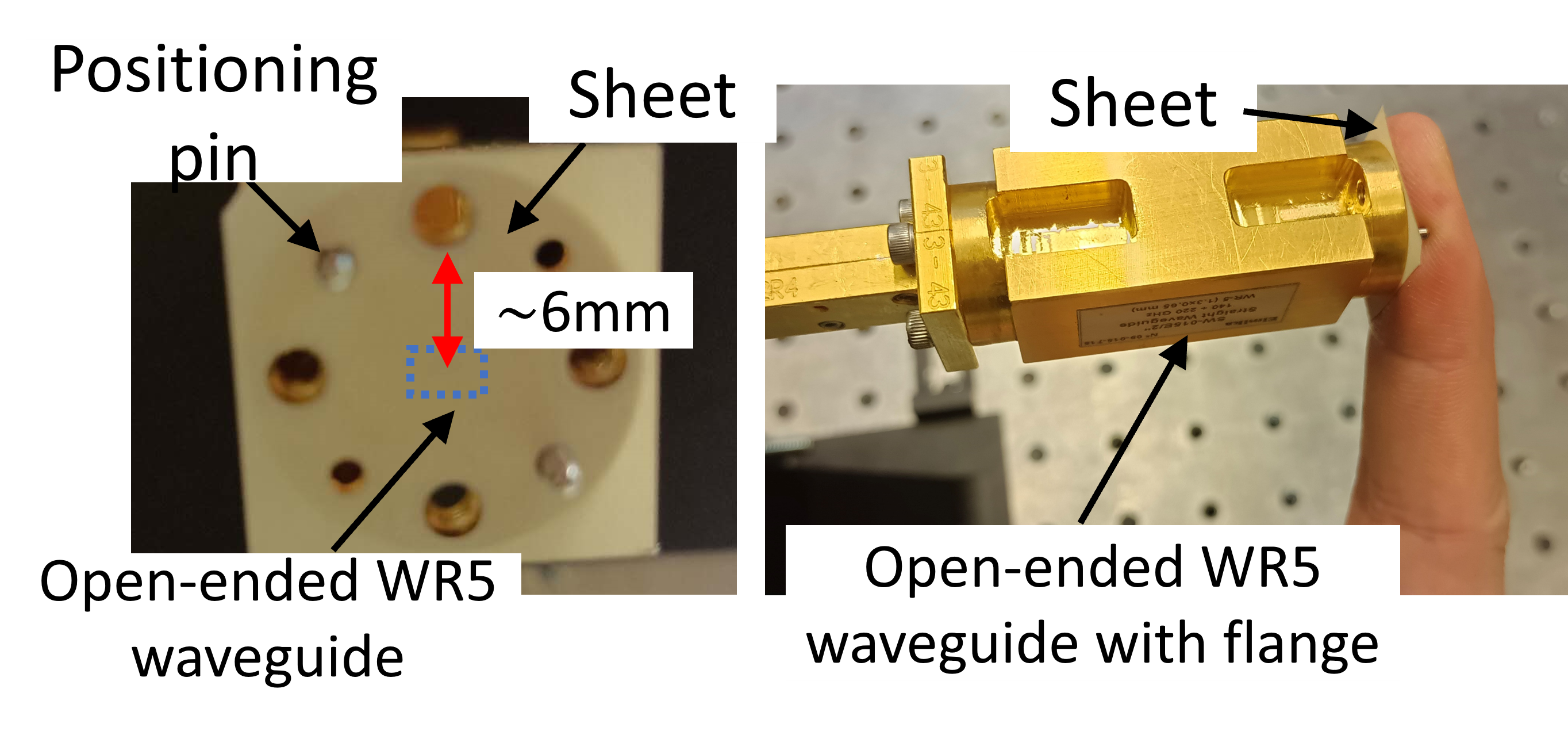}
	\caption{The setup of finger permittivity measurement.}
	\label{fig:permmeasurement}
\end{figure}

The measurement setup is illustrated in Fig.~\ref{fig:permmeasurement}. 
During our measurements, the fingers under consideration were not washed or dried to maintain a typical daily humidity level encountered during mobile terminal usage. In the context of finger-antenna interactions, particularly during mobile terminal usage, it is the fingerprint sides that notably influence the radiation patterns of the mobile terminal antennas. Consequently, the measurements primarily focus on the fingerprint sides, while the fingernail sides are not considered.

Additionally, different parts of the fingers, especially the fingerprint regions, may display varying permittivities. Hence, averaging permittivity estimates from multiple finger measurements is essential for obtaining representative values. To collect finger permittivity estimates across various individuals, we conducted measurements with 10 volunteer subjects aged between 18 and 40 years old, and their heights are from 1.60 to 1.85 meters. Each volunteer underwent 10 measurements, focusing on the fingertips of their index, middle, and ring fingers.

 \section{Results and Discussions}
\begin{figure}[!ht]
 \vspace{-\baselineskip}
    \centering
\subfloat[]{
\centering\label{fig:complexpermit}
     \includegraphics[width=0.5\linewidth]{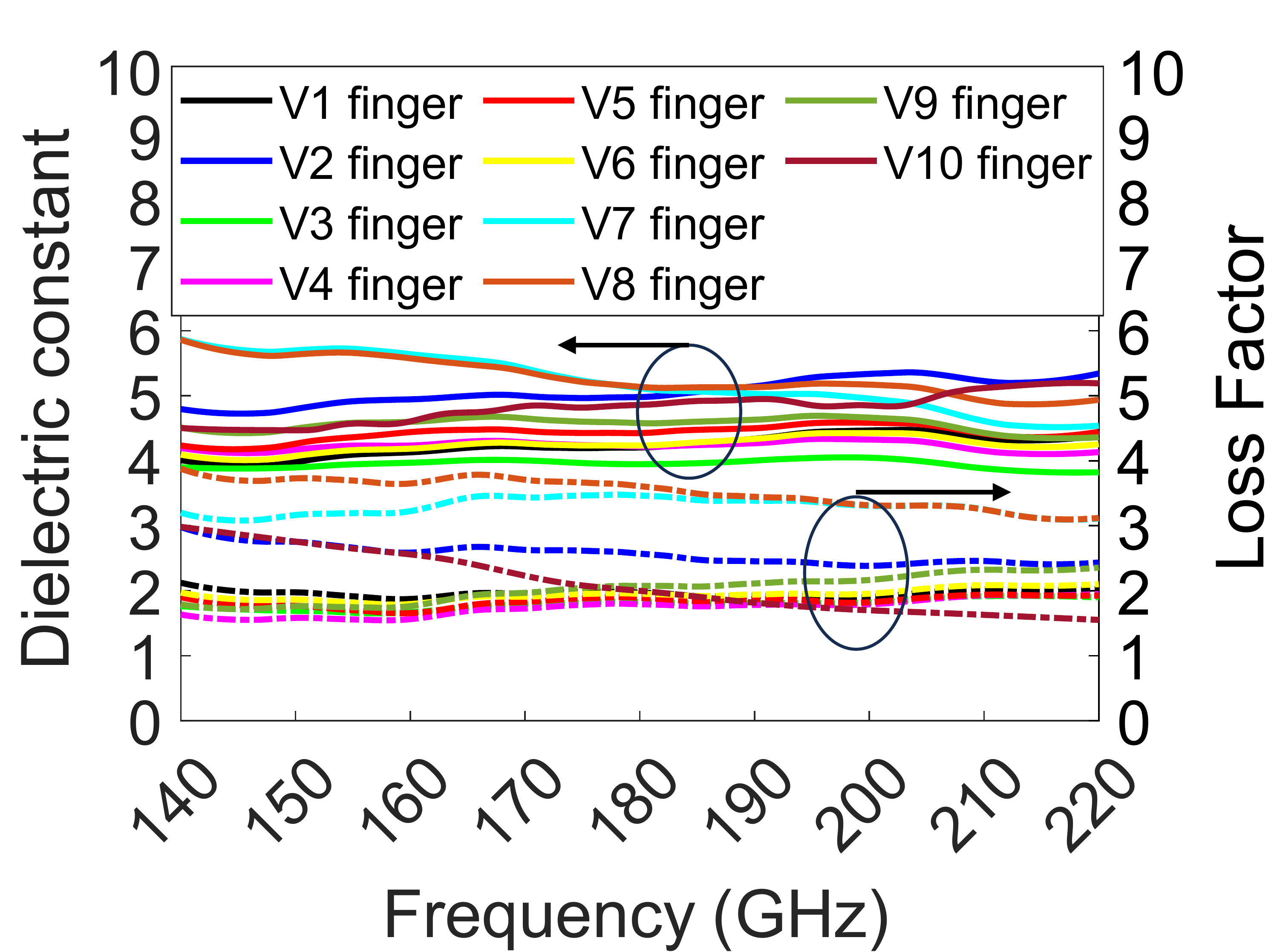}}
\subfloat[]{
\centering
\label{fig:mean}\includegraphics[width=0.5\linewidth]{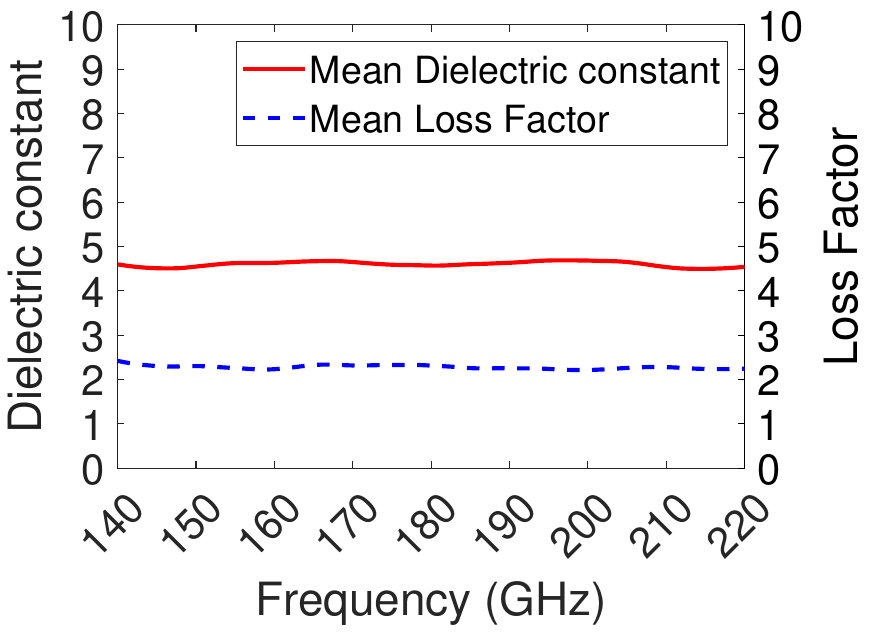}}
	\caption{Mean dielectric constant and loss factor based on 10 different finger parts for a) each volunteer and b) all the volunteers.}
	\label{fig:fgd}
 \vspace{-\baselineskip}
\end{figure}

The mean of the estimated relative permittivities is depicted in Fig.~\ref{fig:complexpermit} for each volunteer. For instance, at $140~\rm GHz$, the dielectric constant (real part) ranges from $4.0$ to $5.9$, while the loss factor (imaginary part) varies from $1.9$ to $3.9$. In most cases, the dielectric constants fall within the range of $3.9$ to $4.8$, while the loss factors fall between $1.9$ to $2.4$. 
The average relative permittivity of all volunteers is $\varepsilon_{\rm r} = 4.7-2.4\rm i$ at $140~\rm GHz$. It can be seen that Volunteer 7 and 8 show higher dielectric constants and loss factors and show different trends with frequency changes. This is because these two volunteers' hands are more moist.
\begin{figure}[!ht]
 \vspace{-\baselineskip}
    \centering
\subfloat[]{
\centering
     \includegraphics[width=0.5\linewidth]{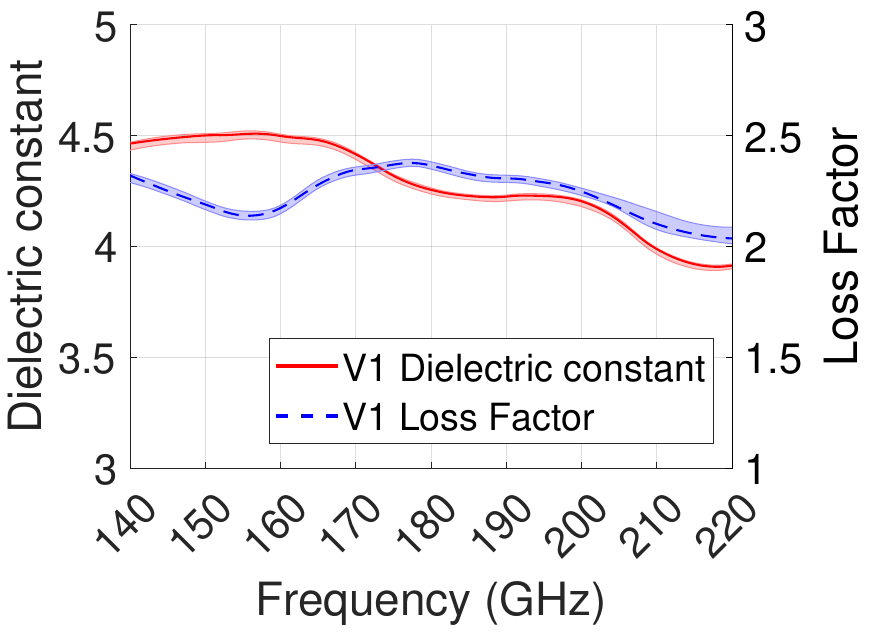}}
\subfloat[]{
\centering
\includegraphics[width=0.5\linewidth]{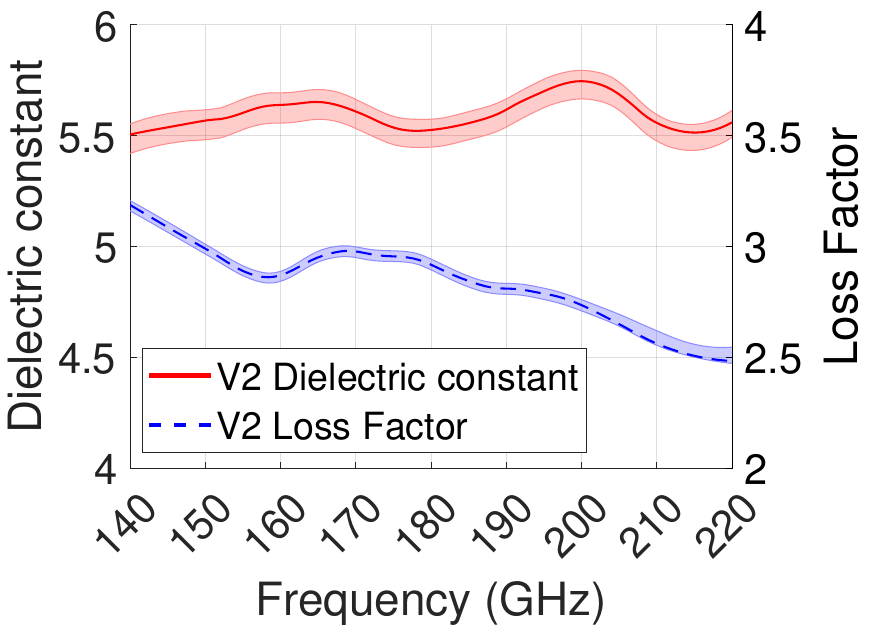}}
	\caption{Measurement uncertainty of dielectric constant and loss factor based on 20 measurements at the same finger location  for a)  Volunteer 1 and b) Volunteer 2.}
	\label{Inhomogeneity}
\end{figure}

\begin{figure*}[!ht]
    \centering
\subfloat[]{
\centering
\label{fig:V1}
     \includegraphics[width=0.195\linewidth]{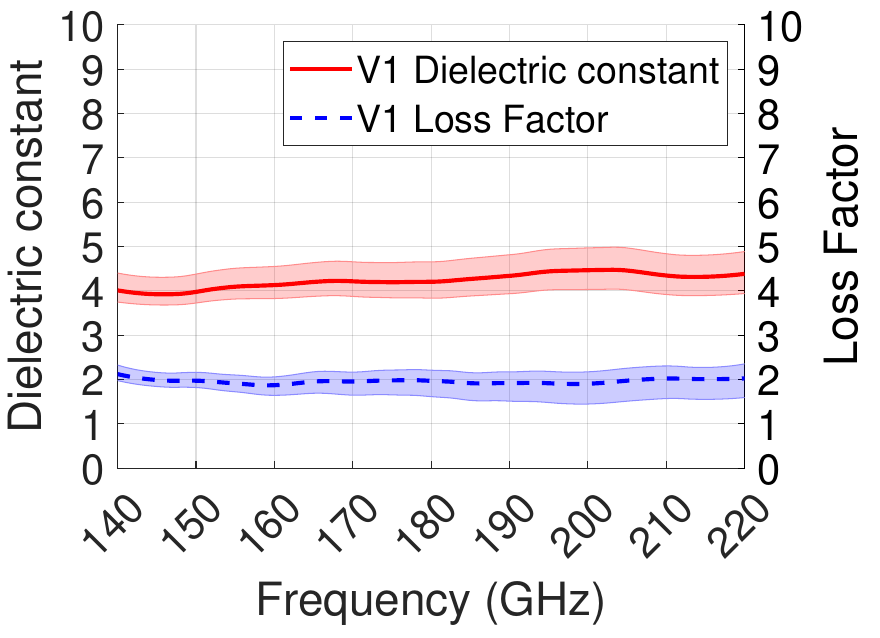}}
\subfloat[]{
\centering
\label{fig:V2}\includegraphics[width=0.195\linewidth]{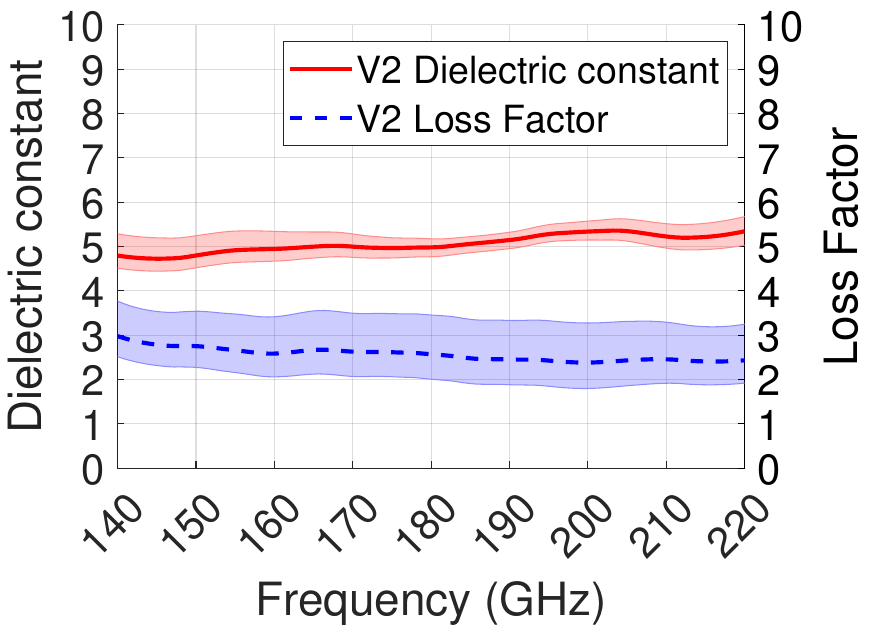}}
\subfloat[]{
\centering
\label{fig:V3}\includegraphics[width=0.195\linewidth]{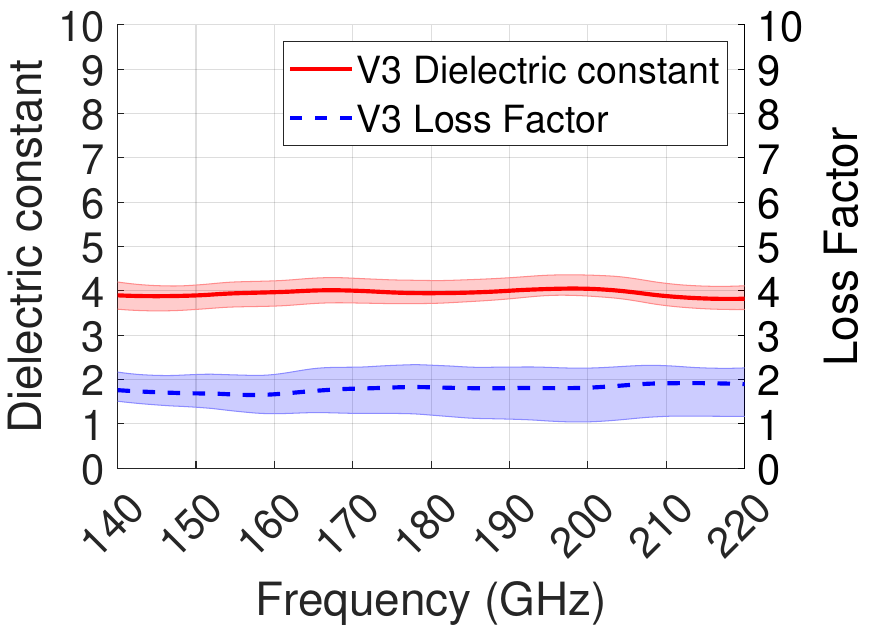}}
\subfloat[]{
\centering
\label{fig:V4}\includegraphics[width=0.195\linewidth]{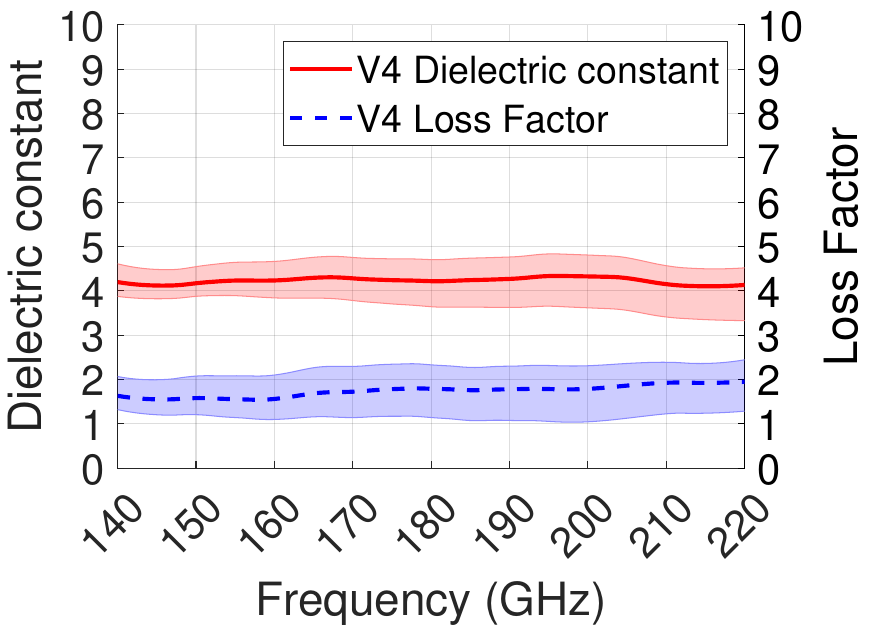}}
\subfloat[]{
\centering
\label{fig:V5}\includegraphics[width=0.195\linewidth]{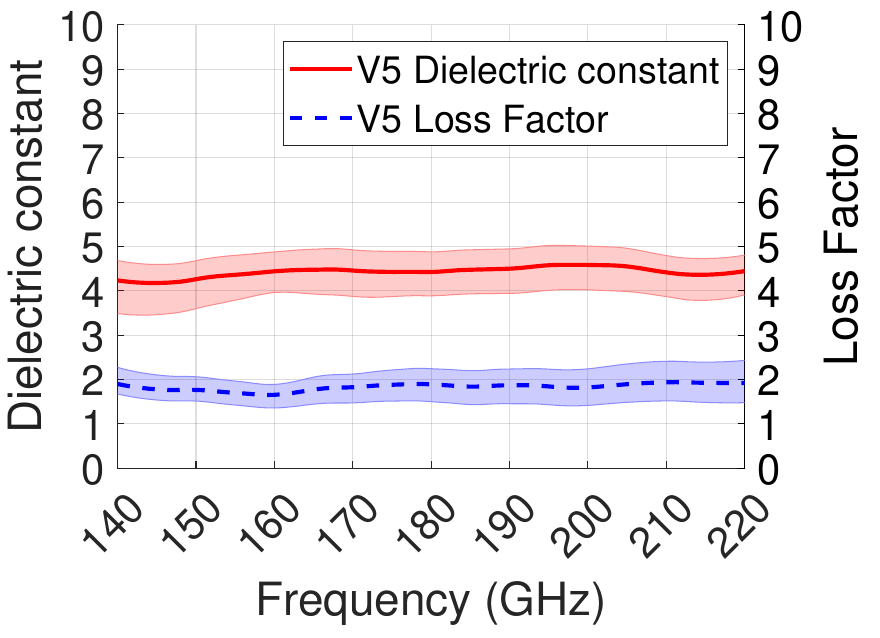}}
\\
\subfloat[]{
\centering
\label{fig:V6}\includegraphics[width=0.195\linewidth]{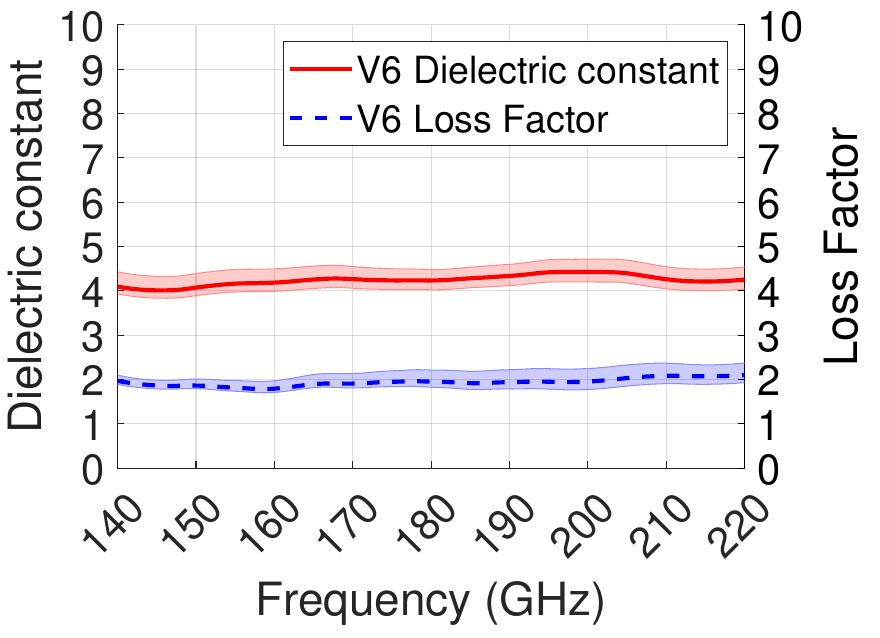}}
\subfloat[]{
\centering
\label{fig:V7}\includegraphics[width=0.195\linewidth]{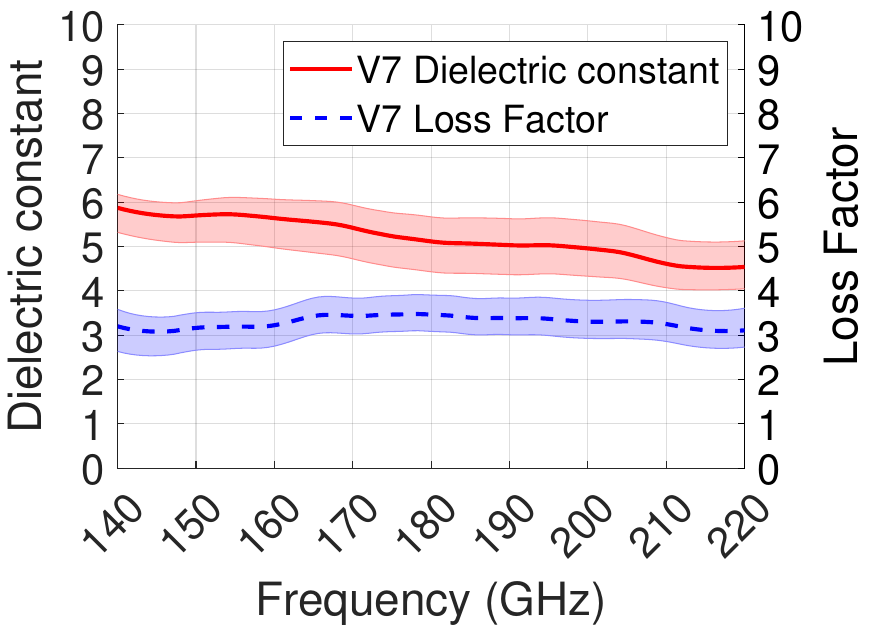}}
\subfloat[]{
\centering
\label{fig:V8}\includegraphics[width=0.195\linewidth]{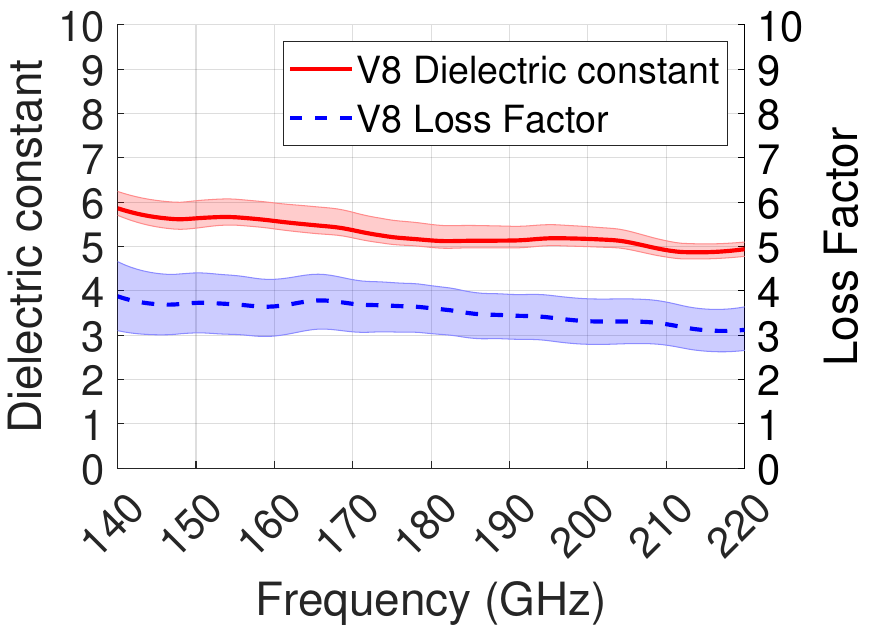}}
\subfloat[]{
\centering
\label{fig:V9}\includegraphics[width=0.195\linewidth]{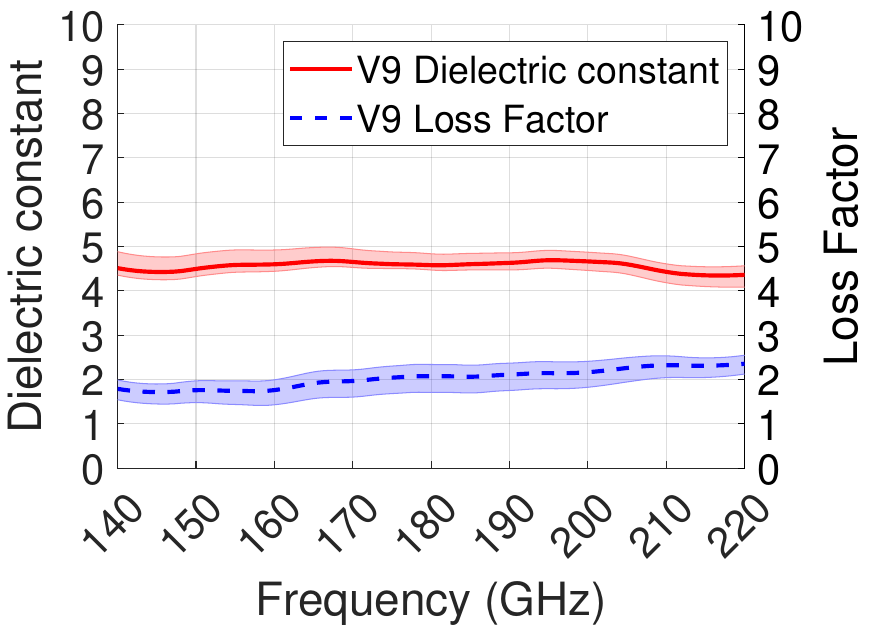}}
\subfloat[]{
\centering
\label{fig:V10}\includegraphics[width=0.195\linewidth]{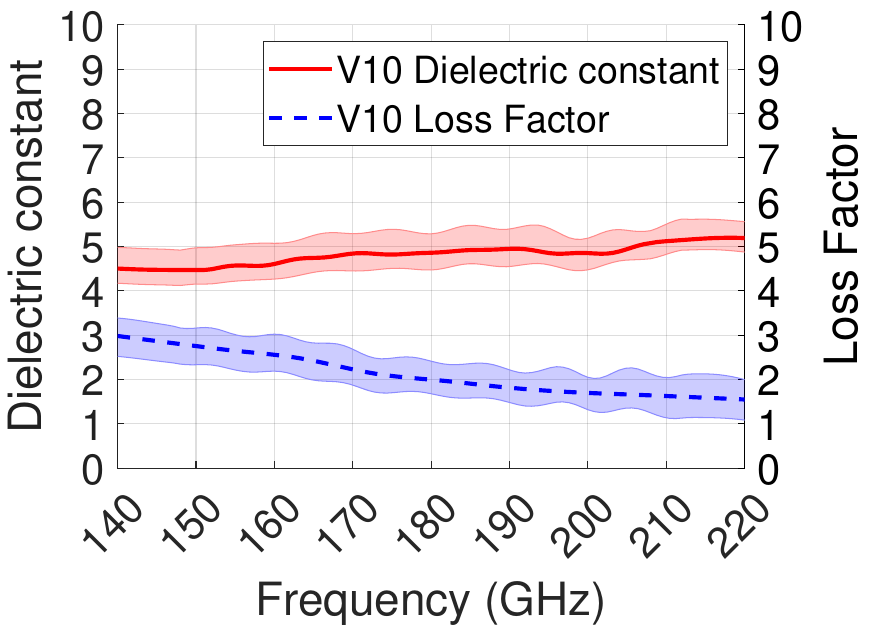}}
	\caption{(a)-(j) Dielectric constant and loss factor based on 10 different finger parts for 10 volunteers.}
	\label{fig:persons}
 \vspace{-\baselineskip}
\end{figure*}
To assess measurement uncertainty, 20 repeated measurements were conducted within a short duration of 2 minutes at the same finger location of two volunteers. The estimated relative deviation of permittivity is within $\pm 1.4\%$ of the mean value for the dielectric constants and within $\pm 1.2\%$ for loss factors, as presented in Fig.~\ref{Inhomogeneity}. These relative deviations are smaller than those reported in~\cite{gao2018towards}, indicating the high repeatability of the measurement method. Additionally, a variation in permittivity for each volunteer was observed based on the 10 measurements on different finger parts shown in Fig.~\ref{fig:persons}. The majority of volunteers' fingers exhibit a variation of $<\pm 0.5$ for both the dielectric constants and loss factors. Some volunteers' dielectric constants show a larger variation, like Fig.~\ref{fig:V4},~\ref{fig:V5}, and~\ref{fig:V7}, while some of the others show a larger variation in loss factors, like Fig.~\ref{fig:V2},~\ref{fig:V3}, and~\ref{fig:V8}. 

\section{Conclusions}
\label{sec:conclusion}
In this manuscript, we propose a novel method for characterizing human skin permittivity using an open-ended waveguide at sub-THz frequencies. The necessary contact area of the human skin under test on the sheet is identified through full-wave simulations so that the antenna-hand interaction model can excellently emulate practical measurements. The full-wave simulations provide reflection coefficients to train the RBN, which is applied to characterize finger permittivities of living humans. The estimated relative permittivity shows smaller uncertainty than $\pm 1.5\%$ for both the dielectric constants and loss factors, demonstrating the high repeatability of the proposed approach. The measured complex relative permittivity estimates are valuable base for studying finger-antenna interactions of sub-THz mobile handsets. Furthermore, the proposed method can be applied to characterize the permittivity of various parts of human skin.

\section*{Acknowledgments}
The authors thank those volunteers engaged in finger-permittivity measurements at the School of Electrical Engineering of Aalto University.

The results presented in this paper have been supported by the Academy of Finland -- NSF joint call pilot ``Artificial intelligence and wireless communication technologies", decision \# 345178.

\printbibliography

@STRING{IEEE_J_MTT        = "{IEEE} Trans. Microw. Theory Techn."}

@STRING{IEEE_J_TTHZ       = "{IEEE} Trans. {THz} Sci. Technol."}

@STRING{IEEE_J_IM         = "{IEEE} Trans. Instrum. Meas."}

@STRING{IEEE_O_ACC        = "{IEEE} Access"}

@STRING{IEEE_M_COM        = "{IEEE} Commun. Mag."}

@article{hosseini2016wideband,
  title={Wideband nondestructive measurement of complex permittivity and permeability using coupled coaxial probes},
  author={Hosseini, Mohammad Hossein and Heidar, Hamid and Shams, Mohammad Hossein},
  journal=IEEE_J_IM,
  volume={66},
  number={1},
  pages={148--157},
  year={2016},
  publisher={IEEE}
}

@article{wang2022fast,
  title={Fast and robust characterization of lossy dielectric slabs using rectangular waveguides},
  author={Wang, Xuchen and Tretyakov, Sergei A},
  journal=IEEE_J_MTT,
  volume={70},
  number={4},
  pages={2341--2350},
  year={2022},
  publisher={IEEE}
}

@inproceedings{aliouane2022material,
  title={Material Reflection Measurements in Centimeter and Millimeter Wave ranges for {6G} Wireless Communications},
  author={Aliouane, Mohamed Abdelbasset and Conrat, Jean-Marc and Cousin, Jean-Christophe and Begaud, Xavier},
  booktitle={{2022 Joint European Conference on Networks and Communications \& 6G Summit (EuCNC/6G Summit)}},
  pages={43--48},
  year={2022},
  organization={IEEE}
}

@article{gao2018towards,
  title={Towards accurate and wideband in vivo measurement of skin dielectric properties},
  author={Gao, Yuan and Ghasr, Mohammad Tayeb and Nacy, Michael and Zoughi, Reza},
  journal=IEEE_J_IM,
  volume={68},
  number={2},
  pages={512--524},
  year={2018},
  publisher={IEEE}
}

@ARTICLE{Zhu2021complex,
  author={Zhu, Hao-Tian and Wu, Ke},
  journal=IEEE_J_TTHZ, 
  title={Complex Permittivity Measurement of Dielectric Substrate in {Sub-THz} Range}, 
  year={2021},
  volume={11},
  number={1},
  pages={2-15},
  doi={10.1109/TTHZ.2020.3036181}}

@ARTICLE{Krogerus2007,
  author={Krogerus, Joonas and Toivanen, Juha and Icheln, Clemens and Vainikainen, Pertti},
  journal=IEEE_J_IM, 
  title={Effect of the Human Body on Total Radiated Power and the 3-{D} Radiation Pattern of Mobile Handsets}, 
  year={2007},
  month=Dec,
  volume={56},
  number={6},
  pages={2375-2385},
  doi={10.1109/TIM.2007.903591}}

@book{pozar2011,
  title={Microwave engineering},
  author={Pozar, David M},
  year={2011},
  publisher={John wiley \& sons}
}

@ARTICLE{Xue2022,
  author={Xue, Bing and Koivum\"aki, Pasi and V\"ah\"a-Savo, Lauri and Haneda, Katsuyuki and Icheln, Clemens},
  journal=IEEE_J_IM, 
  title={Impacts of Real Hands on 5{G} Millimeter-Wave Cellphone Antennas: Measurements and Electromagnetic Models}, 
  year={2023},
  volume={72},
  number={},
  pages={1-12},
  doi={10.1109/TIM.2023.3267350}}

@article{Zhekov2019,
  author={Zhekov, Stanislav Stefanov and Franek, Ondrej and Pedersen, Gert Frølund},
  journal=IEEE_O_ACC, 
  title={Dielectric Properties of Human Hand Tissue for Handheld Devices Testing}, 
  year={2019},
  volume={7},
  number={},
  pages={61949-61959},
  month=May,
  ISSN={2169-3536},
  doi={10.1109/ACCESS.2019.2914863}}

@article{feldman2009electromagnetic,
  title={The electromagnetic response of human skin in the millimetre and submillimetre wave range},
  author={Feldman, Yuri and Puzenko, Alexander and Ishai, Paul Ben and Caduff, Andreas and Davidovich, Issak and Sakran, Fadi and Agranat, Aharon J},
  journal={Physics in Medicine \& Biology},
  volume={54},
  number={11},
  pages={3341},
  year={2009},
  publisher={IOP Publishing}
}

@article{zakharov2009full,
  title={Full-field optical coherence tomography for the rapid estimation of epidermal thickness: study of patients with diabetes mellitus type 1},
  author={Zakharov, P and Talary, MS and Kolm, I and Caduff, A},
  journal={Physiological measurement},
  volume={31},
  number={2},
  pages={193},
  year={2009},
  publisher={IOP Publishing}
}

@article{sasaki2017monte,
  title={Monte Carlo simulations of skin exposure to electromagnetic field from 10 {GHz} to 1 {THz}},
  author={Sasaki, Kensuke and Mizuno, Maya and Wake, Kanako and Watanabe, Soichi},
  journal={Physics in Medicine \& Biology},
  volume={62},
  number={17},
  pages={6993},
  year={2017},
  publisher={IOP Publishing}
}

@article{betzalel2017modeling,
  title={{The modeling of the absorbance of sub-THz radiation by human skin}},
  author={Betzalel, Noa and Feldman, Yuri and Ishai, Paul Ben},
  journal=IEEE_J_TTHZ,
  volume={7},
  number={5},
  pages={521--528},
  year={2017},
  publisher={IEEE}
}

@article{zhang2022out,
  title={Out-of-Band Information Aided {mmWave/THz} Beam Search: A Spatial Channel Similarity Perspective},
  author={Zhang, Peize and Kyosti, Pekka and Haneda, Katsuyuki and Koivumaki, Pasi and Lyu, Yejian and Fan, Wei},
  journal=IEEE_M_COM,
  year={2022},
  publisher={IEEE}
}

@inproceedings{nor2022effect,
  title={The Effect of Human Blockage on The Performance of {RIS} aided Sub-{THz} Communication System},
  author={Nor, Ahmed M and Fratu, Octavian and Halunga, Simona},
  booktitle={Proc. 14th Int. Conf. Computational Intelligence and Commun. Netw. (CICN 2022)},
  pages={622--625},
  year={2022},
  organization={IEEE}
}

\end{document}